\documentclass[utf8]{frontiersSCNS} 

\setcitestyle{square} 
\usepackage{url,hyperref,lineno,microtype,subcaption}
\usepackage[onehalfspacing]{setspace}
\usepackage{aas_macros}

\def\keyFont{\fontsize{8}{11}\helveticabold }
\def\firstAuthorLast{Joyce Ann Guzik} 
\def\Authors{Joyce Ann Guzik$^{1,*}$}
\def\Address{$^{1}$Los Alamos National Laboratory, Los Alamos, NM 87545  USA
 }
\def\corrAuthor{Joyce Ann Guzik}

\def\corrEmail{joy@lanl.gov}

\begin{document}
\onecolumn
\firstpage{1}

\title[NASA\,\textbf{\textit{Kepler}} $\delta$ Sct Highlights]{Highlights of Discoveries for $\delta$ Scuti Variable Stars from the\,\textbf{\textit{Kepler}} Era} 

\author[\firstAuthorLast ]{\Authors} 
\address{} 
\correspondance{} 

\extraAuth{}

\maketitle

\begin{abstract}


The NASA\,{\textit{Kepler}} and follow-on K2 mission (2009-2018) left a legacy of data and discoveries, finding thousands of exoplanets, and also obtaining high-precision long time-series data for hundreds of thousands of stars, including many types of pulsating variables.  Here we highlight a few of the ongoing discoveries from\,{\textit{Kepler}} data on $\delta$ Scuti pulsating variables, which are core hydrogen-burning stars of about twice the mass of the Sun.  We discuss many unsolved problems surrounding the properties of the variability in these stars, and the progress enabled by\,{\textit{Kepler}} data in using pulsations to infer their interior structure, a field of research known as asteroseismology.

\tiny
 \keyFont{ \section{Keywords:} Stars: $\delta$ Scuti, Stars: $\gamma$ Doradus, NASA Kepler Mission, asteroseismology, stellar pulsation} 
\end{abstract}

\section{Introduction}
The long time-series, high-cadence, high-precision photometric observations of the NASA {\textit{Kepler}} (2009-2013) \citep{2010Sci...327..977B, 2010PASP..122..131G, 2010ApJ...713L..79K} and follow-on K2 (2014-2018) \citep{2014PASP..126..398H} missions have revolutionized the study of stellar variability.  The amount and quality of data provided by\,{\textit{Kepler}} is nearly overwhelming, and will motivate follow-on observations and generate new discoveries for decades to come.

Here we review some highlights of discoveries for $\delta$ Scuti (abbreviated as $\delta$ Sct) variable stars from the\,{\textit{Kepler}} mission.  The $\delta$ Sct variables are pre-main-sequence, main-sequence (core hydrogen-burning), or post-main-sequence (undergoing core contraction after core hydrogen burning, and beginning shell hydrogen burning) stars with spectral types A through mid-F, and masses around 2 solar masses.  They pulsate in one or more radial and nonradial modes with periods of around 2 hours.  The pulsations are driven mainly by the ``$\kappa$-effect'' (opacity-valving) mechanism in the region of the second ionization of helium at temperatures around 50,000 K in the stellar envelope \citep{2010aste.book.....A}.

Several reviews on\,{\textit{Kepler}} findings for $\delta$ Sct variables have already been written (see, e.g., \citet{2018FrASS...5...43B,2018MNRAS.476.3169B}), and a comprehensive review of the\,{\textit{Kepler}} legacy for these stars is premature.  Prior to\,{\textit{Kepler}}, one of the best compilations of the state-of-the-art of research on $\delta$ Sct variables was the Handbook and Conference Proceedings volume {\it $\delta$ Scuti and Related Stars} \citep{2000ASPC..210.....B}.  New catalogs and lists of variable stars, including $\delta$ Sct stars observed and first discovered by the\,{\textit{Kepler}}/K2 missions, have begun to appear (e.g., \citet{2015AJ....149...68B, 2016AJ....151...86B}, 84 $\delta$ Sct and 32 hybrid (see Section \ref{sec:hybrids}) candidates from\,{\textit{Kepler}} Guest Observer Program Cycle 1-5 observations; \citet{2019MNRAS.485.2380M}, 1988 $\delta$ Sct stars from \,{\textit{Kepler}} observations; \citet{2019FrASS...6...40G}, 249 $\delta$ Sct candidates from K2 observations).

After the successes of asteroseismology to infer the interior structure of the Sun and properties of sun-like stars, studying the slightly more massive $\delta$ Sct stars appeared to be a promising next direction for asteroseismology.  Before the space observations of\,{\textit{Kepler}}, CoRoT (see, e.g., \citet{2009A&A...506...85P}), and MOST (see, e.g., \citet{2007CoAst.150..333M}), there existed only around a dozen $\delta$ Sct stars with long time-series observations from ground-based networks allowing the detection of a large number of pulsation modes (e.g., FG Vir \citep{2005A&A...435..955B} or 4 CVn \citep{2017A&A...599A.116B}) that could be used to constrain stellar models.  The field of $\delta$ Sct asteroseismology has been impeded by the problem of mode identification for several reasons.  Unlike for the Sun, the disks of distant stars cannot be highly resolved, so only low-degree ($\ell$ $\lesssim$ 3) mode variations that do not average out over the disk can be detected photometrically.  Furthermore, most of these stars rotate more rapidly than the Sun, resulting in large and uneven rotational splittings, such that multiplets of adjacent modes can overlap.  In addition, not all of the modes expected by nonadiabatic pulsation calculations are found in the observations.  Finally, the modes are of low radial order $n$, and therefore the spacing pattern is not expected to show regular large separations seen for the higher-order stochastically excited solar-like modes where $n\sim$20 and the modes can be described using asymptotic theory ($n \gg l$; see, e.g., \citet{2019LRSP...16....4G}).  Unlike for the Sun, fundamental properties of a single $\delta$ Sct variable (mass, radius, age, detailed element abundances) cannot be derived from complementary or independent observations (e.g., meteorites, Earth or planetary orbits).  $\delta$ Sct stars in clusters, binaries, or having planetary systems are therefore useful to provide additional constraints for modeling.

The A-F main-sequence stars occupy a small region in the center of the H-R diagram, but have been differentiated into not only $\delta$ Sct and $\gamma$ Doradus (abbreviated as $\gamma$ Dor) variables \citep{2009AIPC.1170..455P, 2011MNRAS.415.3531B}, but also metallic-line A (Am), \citep{2017MNRAS.465.2662S}, peculiar abundance A (Ap) \citep{2020A&A...639A..31M}, rapidly-oscillating Ap (roAp) \citep{2021arXiv210212198H} stars, $\alpha$$^2$ CVn variables \citep{2019MNRAS.487.4695S}, $\lambda$ Boo stars \citep{2017MNRAS.466..546M}, the High Amplitude $\delta$ Sct (HADS) \citep{2000ASPC..210..373M} and SX Phe \citep{2017MNRAS.466.1290N} stars, blue stragglers \citep{2021arXiv210306004R}, and the maybe 'mythical'\footnote{This term was coined by \citet{1980SSRv...27..361B} re. reports of pulsating variables between the main-sequence $\delta$ Sct  and Slowly-Pulsating B-type stars in the H-R diagram.  There has been debate in the literature about whether these variables actually exist, what drives their pulsations, and whether they should be considered a separate class of variable stars.  The prototype Maia, a member of the Pleiades, has been shown not to pulsate (see, e.g., \citet{2017MNRAS.471.2882W}).  Other proposed members of the class may be rapidly rotating SPB stars, or be slowly rotating, but have pulsations driven by the $\kappa$ effect with enhanced opacity around 125,000 K \citep{2017EPJWC.16003013D}.} Maia variables \citep{2018FrASS...5...43B,2017MNRAS.471.2882W,2017EPJWC.16003013D,1980SSRv...27..361B,1983apum.conf....1C}.  \textit{Kepler} observations have revealed overlap and commonalities among these types, pointing the way to a more fundamental understanding of the origins of the diverse phenomena seen in these stars.   

While these stars were expected at first to be the next straightforward step beyond the solar-like oscillators for applications of asteroseismology, this goal has turned out to be more difficult to achieve than expected.  However, the many complexities of these stars make this field of variable star research rich in potential discoveries.

\section{Discovery highlights}

\vspace{0.1in}
\subsection{Pre- and post-\textbf{\textit{Kepler}} view—hybrids everywhere!}
\label{sec:hybrids}
\vspace{0.1in}

Before\,{\textit{Kepler}, the $\delta$ Sct and $\gamma$ Dor stars and their hybrids were found in the instability regions expected by theory.  The $\delta$ Sct $p$-mode pulsations are driven by the $\kappa$ mechanism in their radiative envelopes, but the longer period (1-3 day) $\gamma$ Dor $g$-mode pulsations are proposed to be driven by the `convective blocking' mechanism operating at the base of their convective envelope around 300,000 K \citep{2000ApJ...542L..57G}.  Using a time-dependent convection treatment, hybrid stars pulsating in both $p$ and $g$ modes were expected and found in a small region of the H-R diagram where these two instability regions overlapped \citep{2005A&A...435..927D}.

Just after the first\,{\textit{Kepler}} light curves were received, it became apparent that this picture would be shattered.  $\delta$ Sct and $\gamma$ Dor variables and their hybrids were found throughout and even somewhat beyond the edges of the combined instability regions \citep{2010ApJ...713L.192G, 2011A&A...534A.125U} (see Fig. \ref{uytterhoevenfig10}). Low-frequency pulsation modes identified using the long (30-minute) cadence\,{\textit{Kepler}} data could have been mis-identified as $\gamma$ Dor $g$ modes, but actually may be Nyquist reflections of frequencies above 24.5 d$^{-1}$.  However, the low frequencies are also found using short (1-minute) cadence data, and Nyquist reflection frequencies can be distinguished using a long-enough series (near one\,{\textit{Kepler}} orbital period of 375.2 d) of long-cadence data \citep{2013MNRAS.430.2986M}.  Some low frequencies could possibly be caused by rotation/starspots, undetected binary companions, rotational perturbations of higher frequency modes, combination frequencies, a background star or nearby bright star in the field of view contaminating the light curve, or Rossby or Kelvin waves.  However, \citet{2014MNRAS.437.1476B}, using only short-cadence data, ruled out most of these explanations, and arrives at the bold conclusion that ``all $\delta$ Sct stars are essentially $\delta$ Sct/$\gamma$ Dor hybrids.''

On the other hand, there exist examples of $\delta$ Sct stars that do not show $g$-mode pulsations in the\,{\textit{Kepler}} data. \citet{2017ampm.book.....B} comments on Balona's claim, and discusses an example of a ``pure'' $\delta$ Sct star KIC 5617488, which has no low-frequency peaks with S/N $\geq$ 4.  The few low-frequency peaks visible in the amplitude spectrum have amplitude less a few $\mu$mag.  It is possible that $g$ modes with angular degree $l$ $\gtrsim$3 are undetected in many stars photometrically, but may be discovered spectroscopically.  Such modes have been identified in $\gamma$ Dor variables \citep{2013ASPC..479..105P}, but usually are also accompanied by higher amplitude $l$=1 modes.

New pulsation driving mechanisms are being investigated.  For example, \citet{2015MNRAS.452.3073B} find that an opacity increase of about a factor of two near temperatures of 115,000 K (log $T$ = 5.06) in the stellar envelope can result in instability of some low-frequency modes, but this opacity bump also reduces the range of unstable high-frequency modes.  \citet{2018FrASS...5...43B} highlights theoretical and computational work by \citet{2016MNRAS.457.3163X} with a new treatment of time-dependent convection that allows $\delta$ Sct stars to pulsate in low-frequency modes.

\begin{figure}[h!]
\begin{center}
\includegraphics[width=14cm]{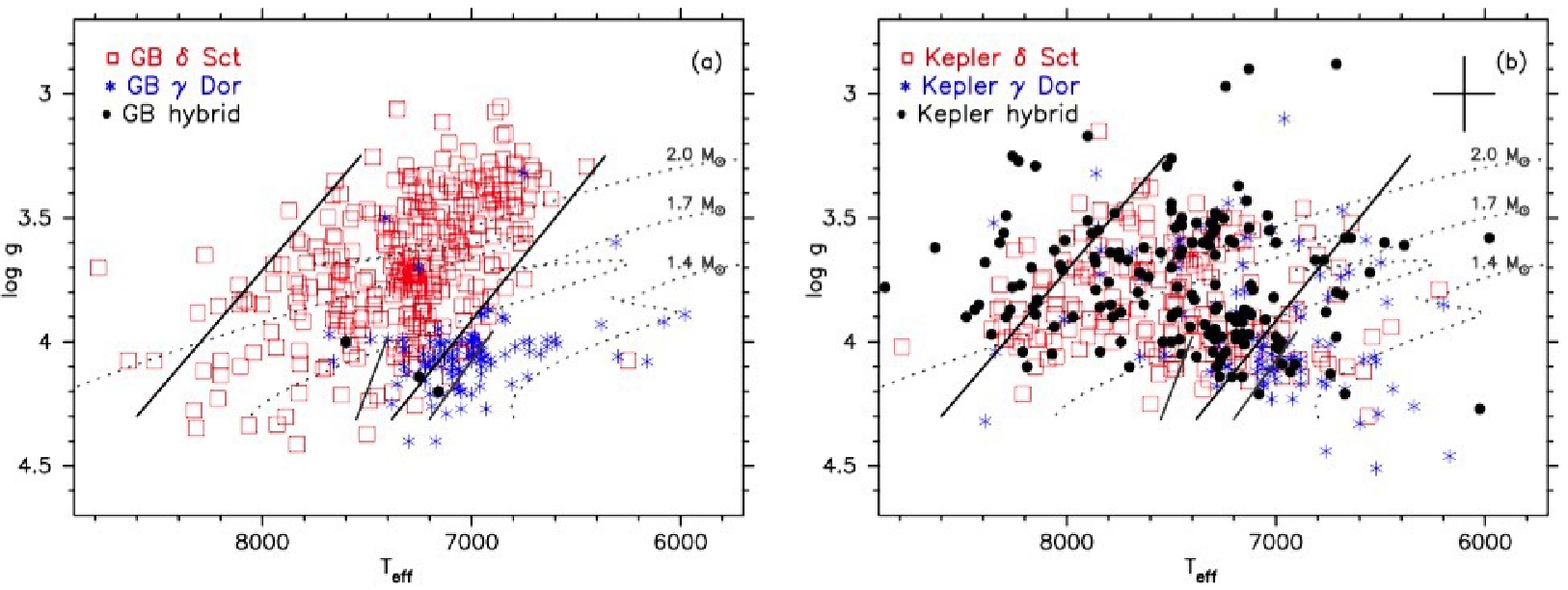}
\end{center}
\caption{Figure 10 from \citet{2011A&A...534A.125U}: a) log surface gravity vs. $T_{\rm eff}$ for the $\delta$ Sct, $\gamma$ Dor, and hybrid stars detected from the ground (parameters taken from the literature). b) log surface gravity vs. $T_{\rm eff}$ for\,{\textit{Kepler}} stars classified as $\delta$ Sct, $\gamma$ Dor, and hybrid stars by \citet{2011A&A...534A.125U}. Open red squares represent $\delta$ Sct stars, blue asterisks indicate $\gamma$ Dor stars, and hybrid stars are marked by black bullets. The black cross in the right top corner shows typical errors on the values. Evolutionary tracks for main-sequence stars with masses 1.4, 1.7, and 2.0 M$_{\odot}$ are plotted with grey dotted lines. The solid thick black and light grey lines mark the blue and red edge of the observed instability strips of $\delta$ Sct and $\gamma$ Dor stars, as described by \citet{2001A&A...366..178R} and \citet{2002MNRAS.333..251H}, respectively.  Reproduced with permission \copyright~ESO.} \label{uytterhoevenfig10}
\end{figure}

\vspace{0.1in}
\subsection{The ``superstar'' and a new pulsation driving mechanism}
\vspace{0.1in}

A $\delta$ Sct star that attracted early excitement was HD 187547 (KIC 7548479), known as the ‘superstar’, observed by\,{\textit{Kepler}} in short cadence.  This star shows not only the expected $\delta$ Sct pulsation modes, but also some additional modes of somewhat higher frequency superimposed (Fig. \ref{antocifig1}).  \citet{2011Natur.477..570A} suggested that convection was stochastically exciting these modes, despite the fact that $\delta$ Sct star models do not have large efficient envelope convection zones, making this star the first $\delta$ Sct/solar-like oscillator discovered.  However, continued\,{\textit{Kepler}} observations showed that the mode lifetimes were quite long, longer than 960 days, and may in fact be ‘coherent’, i.e., not stochastically excited.  \citet{2014ApJ...796..118A} proposed a new pulsation driving mechanism for these higher frequency modes, the `turbulent pressure' mechanism, operating in the outer convective layers of these stars.  They illustrated this mechanism using models including a time-dependent convection treatment applied to radial modes.  This discovery was one of several from the\,{\textit{Kepler}} data resulting in a suggested new pulsation driving mechanism.

\begin{figure}[h!]
\begin{center}
\includegraphics[width=12cm]{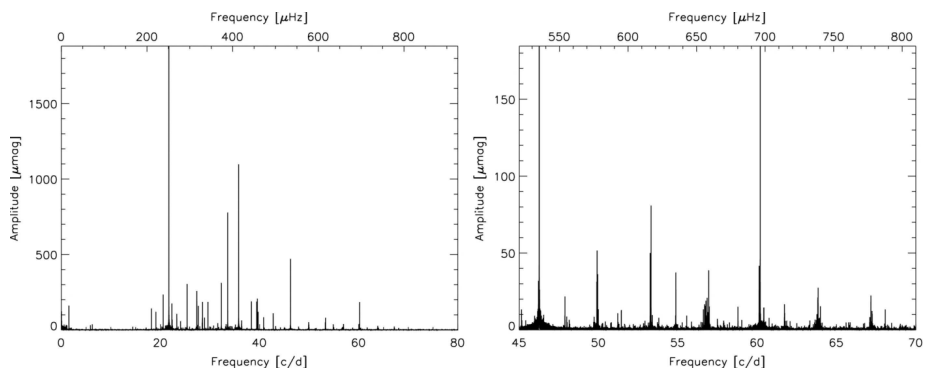}
\end{center}
\caption{Figure 1 from \citet{2014ApJ...796..118A}.  Left panel: Fourier spectra of the\,{\textit{Kepler}} short-cadence data. Right panel: close-up in the frequency region interpreted by \citet{2011Natur.477..570A} to be stochastically excited.  \copyright~AAS. Reproduced with permission.} \label{antocifig1}
\end{figure}

\vspace{0.1in}
\subsection{`Constant' stars in the $\delta$ Sct instability region}
\vspace{0.1in}

While the\,{\textit{Kepler}} data confused the picture of the instability regions for $\delta$ Sct and $\gamma$ Dor stars and their hybrids, these data also affirmed that many of the stars in the $\delta$ Sct instability region of the H-R diagram are ‘constant’, i.e., not pulsating, at least not at levels detectable by\,{\textit{Kepler}} \citep{2014AstRv...9a..41G, 2013AstRv...8c..83G, 2015arXiv150200175G, 2015AstRv..11....1G, 2015MNRAS.452.3073B, 2015MNRAS.447.3948M}.  

\citet{2015MNRAS.452.3073B} found that 1165 out of 2839 stars (41\%) in the $\delta$ Sct temperature region are not pulsating according to\,{\textit{Kepler}} photometry.  \citet{2019MNRAS.485.2380M} use Gaia DR2 \citep{2018A&A...616A...1G} data to derive luminosities and to investigate the pulsator fraction in the instability strip as a function of effective temperature and luminosity, finding that the pulsator fraction peaks at around 70\%  in the middle of the instability strip.

 \citet{2014AstRv...9a..41G, 2013AstRv...8c..83G, 2015arXiv150200175G, 2015AstRv..11....1G} studied two collections (633 and 2100+ stars respectively) of mostly faint stars in the original\,{\textit{Kepler}} field, using long-cadence observations requested to search for $\delta$ Sct and $\gamma$ Dor candidates.  They find many constant stars, showing no variability at the 20 ppm level for frequencies between 0.2 and 24.5 d$^{-1}$.  Most are outside the $\gamma$ Dor and $\delta$ Sct instability regions, but they find six stars in their sample for Quarters 6-13 \citep{2014AstRv...9a..41G, 2013AstRv...8c..83G}, and 15-52 stars, depending on the uncertainty and systematic errors adopted for the\,{\textit{Kepler}} Input Catalog effective temperature and surface gravity, for Quarters 14-17 \citep{2015arXiv150200175G, 2015AstRv..11....1G} that lie within the pulsation instability regions (see Fig. \ref{guzik2015fig1}).

\citet{2015MNRAS.447.3948M} use high-resolution spectroscopy to investigate constant stars (defined as showing no $\delta$ Sct $p$-mode variations above 50 $\mu$mag amplitude) within the $\delta$ Sct instability strip.  They find that most of these stars have peculiar element abundances with enhancements and deficiencies of certain elements compared to solar abundances, and are classified as metallic-line A (Am) stars.  The diffusive settling and radiative levitation believed to cause the abundance anomalies in Am stars would also be expected to deplete helium from the $\delta$ Sct pulsation driving region, and could explain why these stars are not pulsating.  Setting aside the chemically-peculiar stars, \citet{2015MNRAS.447.3948M} find that the remaining stars not pulsating in $\delta$ Sct $p$ modes are near the edges of the instability regions (Fig. \ref{murphy2015fig2}). \citet{2015MNRAS.447.3948M} propose that some of these stars may be in undetected binaries, and therefore have inaccurate effective temperatures and actually may lie outside the instability strip; it is also possible that a binary companion could inhibit pulsations.   Additional investigation is needed to determine whether these explanations apply for all of the `constant' stars.  

\begin{figure}[h!]
\begin{center}
\includegraphics[width=8cm]{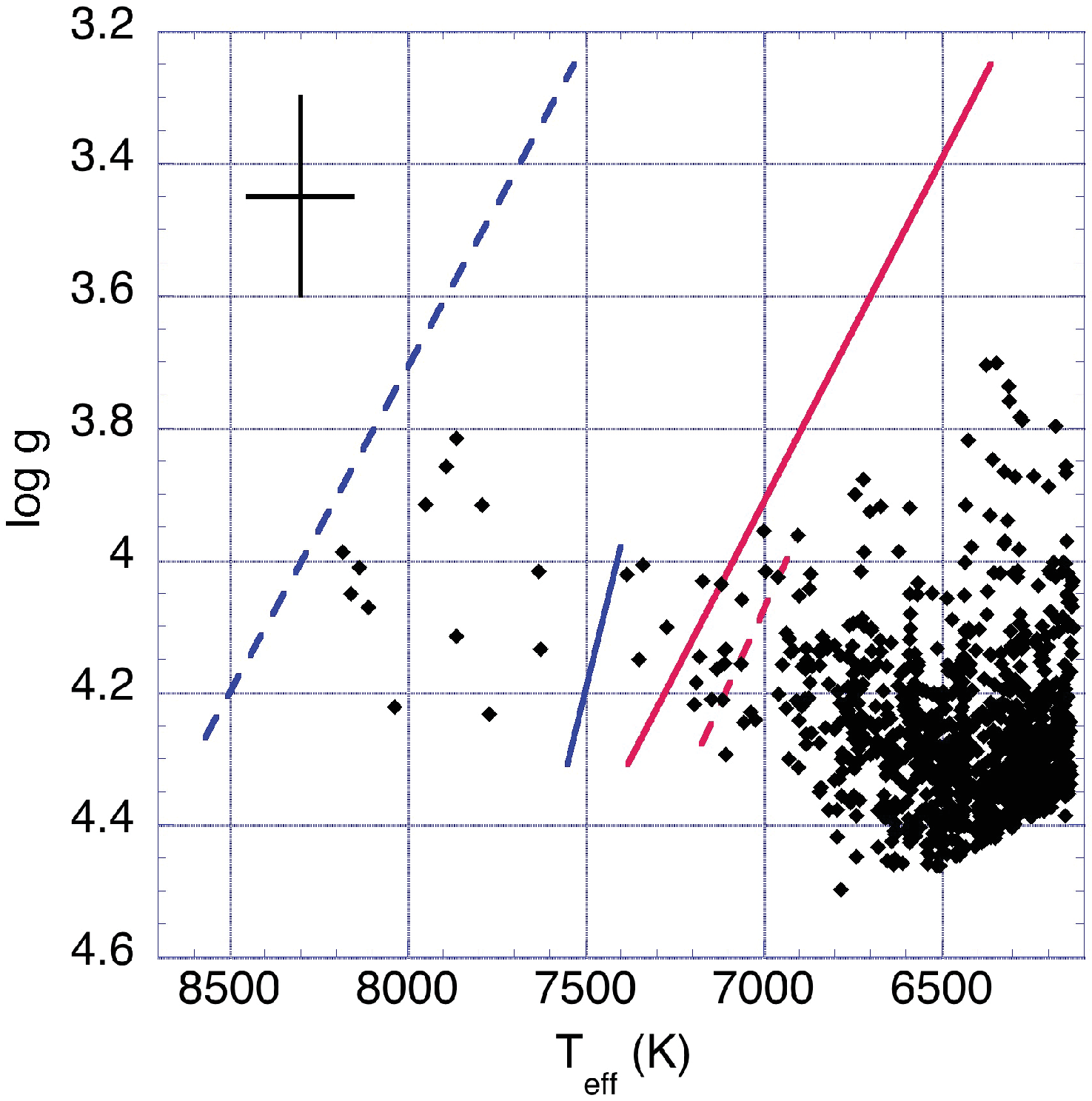}
\end{center}
\caption{Figure 1 from \citet{2015arXiv150200175G}:  Location of stars that are `constant' in the log surface gravity-$T_{\rm eff}$ diagram, along with $\delta$ Sct (dashed lines) and $\gamma$ Dor (solid lines) instability strip boundaries established from pre-\,{\textit{Kepler}} ground-based observations \citep{2001A&A...366..178R, 2002MNRAS.333..251H}.  The $T_{\rm eff}$ of the sample stars has been shifted by +229 K to account for the systematic offset between $T_{\rm eff}$ of the\,{\textit{Kepler}} Input Catalog and SDSS photometry for this temperature range as determined by \citet{2012ApJS..199...30P,2013ApJS..208...12P}. The black cross shows an error bar on log g (0.3 dex) and $T_{\rm eff}$ (290 K) established by comparisons of KIC values and values derived from ground-based spectroscopy for brighter\,{\textit{Kepler}} targets \citep{2011A&A...534A.125U}.  In this figure, 34 `constant' stars lie within the instability strip boundaries.  Without the +229 K offset, 17 'constant' stars would fall within the instability strip boundaries. } \label{guzik2015fig1}
\end{figure}

\begin{figure}[h!]
\begin{center}
\includegraphics[width=10cm]{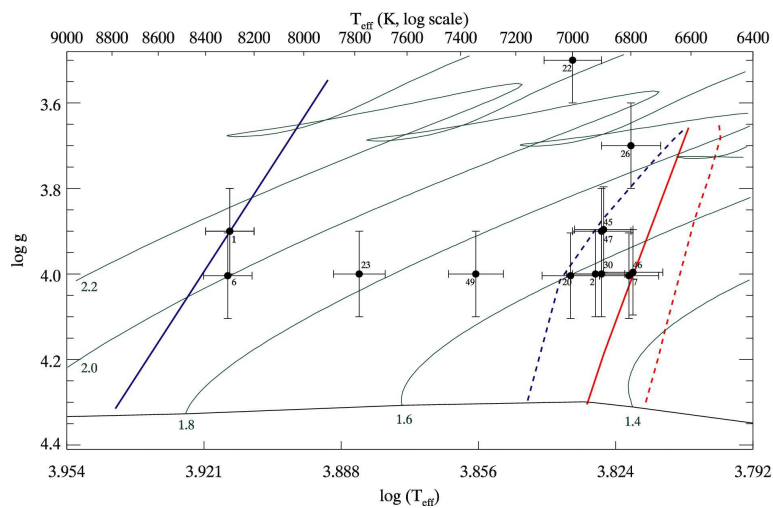}
\end{center}
\caption{Figure 2 from \citet{2015MNRAS.447.3948M}.  Positions of chemically normal, non-$\delta$ Sct stars with 1$\sigma$ error bars. Solid blue and red lines are the blue and red edges of the $\delta$ Sct instability strip, while dashed lines indicate the $\gamma$ Dor instability strip. Green lines are evolutionary tracks, with masses in M$_{\odot}$ written beneath the ZAMS (black). The non-pulsators generally lie near the edges of the $\delta$ Sct instability strip, with exceptions discussed in \citet{2015MNRAS.447.3948M}.} \label{murphy2015fig2}
\end{figure}

\vspace{0.1in}
\subsection{Spots and flares}
\vspace{0.1in}
\citet{2012MNRAS.423.3420B, 2013MNRAS.431.2240B, 2015MNRAS.447.2714B, 2017MNRAS.467.1830B, 2019MNRAS.490.2112B} found that around 40\% of A-type stars observed by\,{\textit{Kepler}}, including many $\delta$ Sct stars, show modulations in their light curves attributed to magnetic activity and starspots, and 1.5\% even show flares.  This behavior is surprising, because hotter stars, including A-type stars, are believed to have thin and inefficient envelope convection layers instead of a larger convective envelope as found in the Sun and cooler stars, and so are not expected to have convection+differential-rotation driven dynamos and magnetic cycles as found in solar-like stars.  \citet{2019MNRAS.490.2112B} published a paper titled ``Evidence for spots on hot stars suggests major revision of stellar physics,'' conveying the significance of these findings.  It is possible that these stars retained a fossil field from their formation.  It is also possible that a dynamo mechanism is operating in the convective core, if a way can be found for the field to diffuse through the overlying radiative layers quickly enough to reach the stellar surface \citep{2005ApJ...629..461B, 2009ApJ...705.1000F}.

Further investigations into A-type flaring stars have been conducted by \citet{2017MNRAS.466.3060P}.  They performed new analyses of the photometry of  33 flaring A-type stars listed by \citet{2012MNRAS.423.3420B, 2013MNRAS.431.2240B}, verifying flares in 27 of these objects.  In fourteen cases, an overlapping object in the\,{\textit{Kepler}} pixel data may be responsible for the flares; in five other cases, the light curves are contaminated by nearby objects in the field.  They also obtained new high-resolution spectroscopic observations of 22 of these stars, finding that eleven are spectroscopic binary systems, so that an unresolved low-mass companion may actually be producing the flares.  Therefore, they have found possible alternative explanations for all but nine of these stars, six of them without high-resolution spectroscopy, casting some doubt on the A-star flare hypothesis.

Concerning the  A-type stars, \citet{2013MNRAS.431.2240B, 2017MNRAS.467.1830B} attributed a broad unresolved hump of peaks with a higher amplitude sharp peak at the higher frequency edge in \,{\textit{Kepler}} amplitude spectra to multiple star spots with finite lifetimes and differential rotation.  This broad hump has been explained by \citet{2018MNRAS.474.2774S} as r modes (global Rossby waves).   \citet{2018MNRAS.474.2774S} suggest that the resolved higher amplitude peak, usually accompanied by 
a few smaller peaks, is produced by one or a few long-lived star spots that could emerge in weakly magnetic A-type stars.  

\vspace{0.1in}
\subsection{Chemically peculiar stars}
\vspace{0.1in}

The chemically peculiar A-type stars further challenge stellar pulsation and evolution theory.  Diffusive settling of helium from the pulsation-driving region is expected to turn off the $\kappa$-effect mechanism and $\delta$ Sct pulsations in Am stars.  However, some Am stars are observed to pulsate in $\delta$ Sct modes \citep{2020MNRAS.498.4272M, Guzik2020data}. Figure \ref{murphy2020fig10} from \citet{2020MNRAS.498.4272M} shows the location of many Am stars including $\delta$ Sct pulsators, along with the blue edge of the $\delta$ Sct instability region calculated including diffusive settling of helium from the driving region. \citet{2020MNRAS.498.4272M} find that pulsation driving from a Rosseland mean opacity bump at 50,000 K caused by the discontinuous H-ionization edge in bound-free opacity explains the observation of $\delta$ Sct pulsations in Am stars.  \citet{2017MNRAS.465.2662S} propose that $\delta$ Sct pulsations in Am stars are driven by the turbulent-pressure mechanism.  

\citet{2011MNRAS.414..792B} find that the observed location of pulsating Am stars in the H-R diagram does not agree with the location predicted from diffusion calculations. \citet{2015MNRAS.448.1378B} state: ``The fact that so many Am stars are $\delta$ Sct variables is also at odds with the prediction of diffusion theory,'' and even suggest that accretion could be the origin of the metal enhancements.

\citet{2018A&A...616A..77B} review magnetic chemically peculiar A-type (Ap) stars, including those that pulsate, observed during the\,{\textit{Kepler}} K2 mission.  \citet{2018MNRAS.478.2777B} use spectro-polarimetry to detect large-scale kilogauss magnetic fields in several chemically peculiar stars observed during the K2 mission.  In Ap stars,``chemical spots'' form at the magnetic poles that cause brightness contrasts that show up as light curve variations as the star rotates; these variables are also called $\alpha$$^2$} CVn variables.  Misalignment of the dipole magnetic field axis and rotation axis is the preferred explanation for properties of high-frequency $p$ modes of the rapidly-oscillating Ap (roAp) stars (see \citet{1982MNRAS.200..807K} and review by \citet{2021arXiv210212198H}, this collection). 

Strong magnetic fields, as are found in the Ap stars, are expected to suppress the low-overtone pulsations found in $\delta$ Sct stars.  However, \citet{2020MNRAS.498.4272M}, using\,{\textit{Kepler}}\,data, report the first $\delta$ Sct-roAp hybrid, KIC 11296437, having mean magnetic field modulus of 2.8 $\pm$ 0.5 kilogauss, and estimated polar magnetic field strength of 3.0 - 5.2 kilogauss.  Figure \ref{murphy2020fig10} shows the location of this star on the H-R diagram based on asteroseismic models.

\begin{figure}[h!]
\begin{center}
\includegraphics[width=10cm]{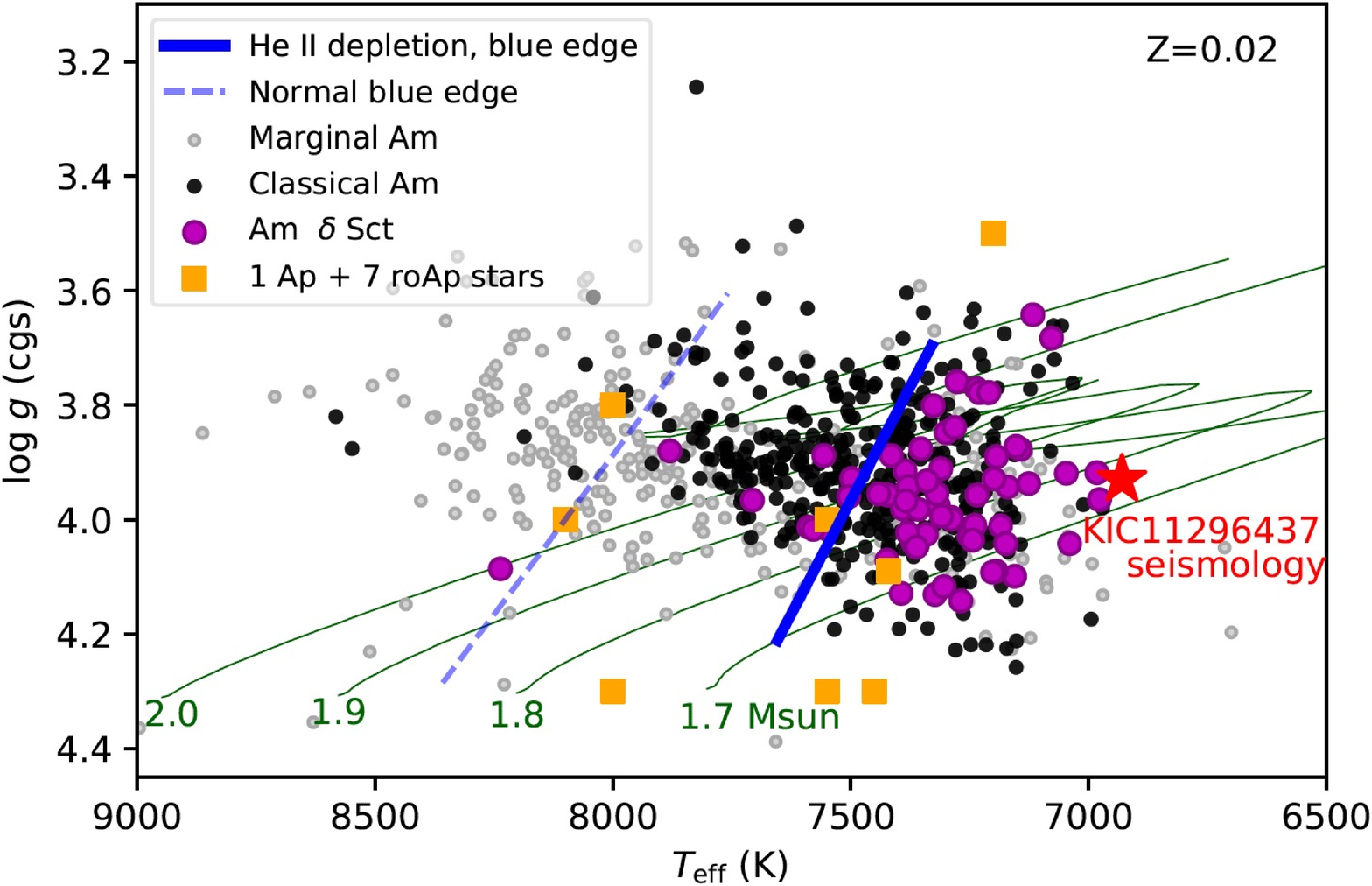}
\end{center}
\caption{Figure 10 from \citet{2020MNRAS.498.4272M}.  The solid blue line marks the fundamental-mode blue edge for stars depleted of helium to the second He ionization zone.  Similar depletion is expected from gravitational settling in Am stars, which are shown according to their degree of peculiarity (from \citet{2017MNRAS.465.2662S}). The pulsating Am stars are highlighted as magenta circles. Evolutionary tracks of helium depleted models with Z=0.02 are shown, with their masses written beneath the ZAMS. Orange squares show the positions of 7 comparison roAp stars, and the post-main-sequence $\delta$ Sct star with Ap-like abundances. The red star is the seismic result for KIC11296437.}  \label{murphy2020fig10}
\end{figure}

\vspace{0.1in}
\subsection{HADS and SX Phe stars}
\vspace{0.1in}

Other A to mid-F spectral type pulsators are further divided into the high-amplitude $\delta$ Sct stars (HADS), and the related SX Phe stars.  The\,{\textit{Kepler}} data show that there may not be any physical distinction between SX Phe and HADS, or really between HADS and normal $\delta$ Sct stars.

\begin{figure}[h!]
\begin{center}
\includegraphics[width=10cm]{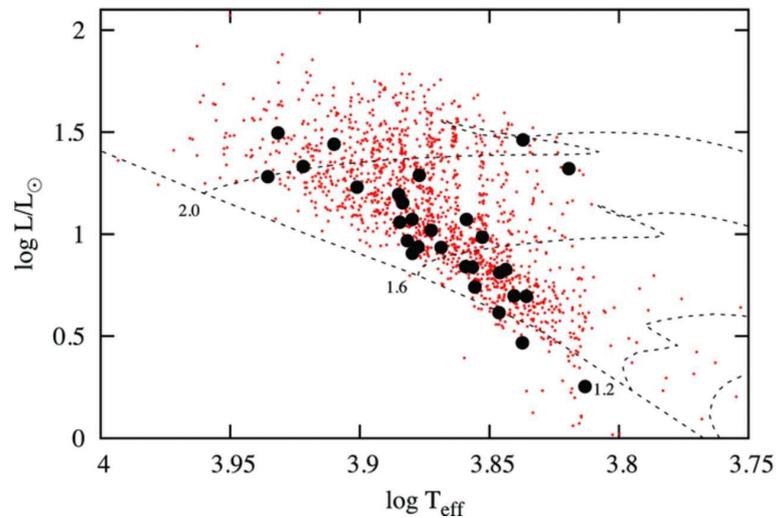}
\end{center}
\caption{Figure 2 from \citet{2012MNRAS.426.2413B}.  The theoretical H-R diagram for\,{\textit{Kepler}} $\delta$ Sct stars (small filled circles). The large filled circles are $\delta$ Sct stars which have large proper motions and large tangential velocities (i.e., SX Phe candidates). Evolutionary tracks are shown and labelled with the solar mass.} \label{balonanemec2012fig2}
\end{figure}

\begin{figure}[h!]
\begin{center}
\includegraphics[width=8cm]{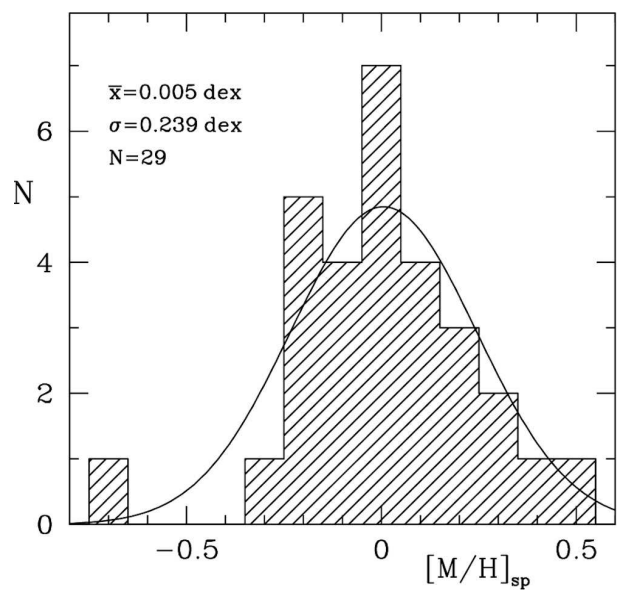}
\end{center}
\caption{Figure 13 from \citet{2017MNRAS.466.1290N}.  Histogram of spectroscopic metal abundances, [M/H]$_{\rm sp}$, for 29 of the \citet{2012MNRAS.426.2413B} SX Phe candidates, fitted with a Gaussian of mean 0.005 dex and standard deviation 0.239 dex; note that KIC 11754974 is off-scale at [M/H]$_{\rm sp}$ = -1.2$\pm$0.3 dex.} \label{nemec2017fig13}
\end{figure}

The SX Phe stars are defined as Population II (low-metallicity) high-amplitude $\delta$ Sct stars, with one or two high-amplitude modes, and are usually found in globular clusters and in dwarf galaxies of the Local Group.  They are bluer and brighter than the cluster turnoffs, and so are called `blue stragglers', which may have been formed by binary mergers.  \citet{2012MNRAS.426.2413B} identified 34 blue straggler candidates in the original\,{\textit{Kepler}} field based on their high tangential velocities (distance $\times$ proper motion), which indicate that they belong to a thick disk or halo population (Fig. \ref{balonanemec2012fig2}).  \citet{2017MNRAS.466.1290N} supplemented the\,{\textit{Kepler}} light curves with new spectroscopic observations to determine metallicity, temperatures, radial velocities, and projected rotational velocity $v$ sin\,$i$.  They found that nearly all of these candidates had near-solar metallicities (Fig. \ref{nemec2017fig13}).  Moreover, the\,{\textit{Kepler}} light curves were not distinguishable from normal $\delta$ Sct stars, as they show complex spectra and even low frequencies as often seen in\,{\textit{Kepler}} $\delta$ Sct light curves.  It may turn out that the defining characteristic of field SX Phe stars, namely showing only one or two high-amplitude modes, is just a selection effect.  There are also low-amplitude multi-periodic SX Phe stars found in globular clusters, lending support to the position that these stars should not be considered a separate class of pulsator from the normal $\delta$ Sct stars.

There is room for debate about whether SX Phe stars should be retained as a class of variables, and if so, how to define them and separate them from HADS or normal $\delta$ Sct stars.   There are questions of cluster vs. field identification, metallicity (Pop. I or II), whether they have undergone mass transfer or merger, whether they are `blue stragglers', i.e., should have evolved off of the main sequence given their metallicity, age, and spectral type.  It may be easiest to identify stars as SX Phe stars if they are found in globular clusters and are blue stragglers located on/near the main sequence above the turnoff.  However, it is more difficult to distinguish them observationally if they are field stars.  As discussed above, for the field stars, a high proper motion does not guarantee metallicity below solar, and so if one were to adopt the criteria of Pop. II metallicity and high proper motion, most of the\,{\textit{Kepler}}  SX Phe candidates would fail this test.  Asteroseismic analysis of the prototype SX Phe \citep{2020svos.conf...81D, 2020MNRAS.499.3034D} shows that it is in the core contraction phase or shell H-burning phase, has age $\sim$ 4 Gyr, and has a low mass (M $\sim$ 1.05 M$_{\odot}$) compared to HADS and $\delta$ Sct stars.  The best-fit models have metallicity Z = 0.0014 to 0.002, and favor a low hydrogen (high helium) abundance X = 0.67 \citep{2020svos.conf...81D}.  One could consider adopting a mass plus age criterion to make the distinction, but these properties are not directly observable and would rely on asteroseismic analyses.  Asteroseismic analyses fortunately show promise to constrain intrinsic metallicity and helium abundance, and possibly whether the star has experienced mass transfer from a binary companion or is the product of a binary merger, either of which could enhance the helium abundance.  Should pulsation mode amplitude or the presence of only one or a few modes be retained as defining criteria?  These questions will require further discussion.  

\vspace{0.1in}
\subsection{`Heartbeat' stars and tidally excited modes}
\vspace{0.1in}

A new variable type categorized as a result of\,{\textit{Kepler}} data are the `heartbeat' stars \citep{2012ApJ...753...86T, 2018MNRAS.473.5165H, 2017MNRAS.472.1538F, 2020ApJ...888...95G, 2020ApJ...903..122C}, with KOI-54 \citep{2011ApJS..197....4W} being the first dramatic example observed by\,{\textit{Kepler}}.  These stars are binaries in highly eccentric ($\epsilon$ $\gtrsim$ 0.3) orbits with orbital periods between 1 day and 1 year that show tidally induced oscillations \citep{2017MNRAS.472.1538F}.  These stars are so-named because their light curves resemble an electrocardiogram, with a brightness dip followed immediately by a sharp rise at periastron.  This feature is caused by increased tidal distortion of the components and viewing them at different angles at they orbit each other at periastron, and is also enhanced by light reflecting from the companion star and Doppler boosting at close approach.  These binaries show tidally excited oscillations that can be identified because they are exact multiples of the binary orbital frequency.  Some have one or more components that also show intrinsic pulsations, and some also show eclipses.  Before\,{\textit{Kepler}}, only a few such systems had been identified, but 17 systems were quickly discovered and characterized from\,{\textit{Kepler}} data \citep{2012ApJ...753...86T}, motivating their grouping as a new class of variables.  It is interesting that \citet{2016AJ....151...68K} catalog 176 heartbeat systems in the original\,{\textit{Kepler}} field, but most have not been closely studied, and only around 20\% actually show the expected tidally excited oscillations \citep{2020ApJ...903..122C}. 

\begin{figure}[h!]
\begin{center}
\includegraphics[width=10cm]{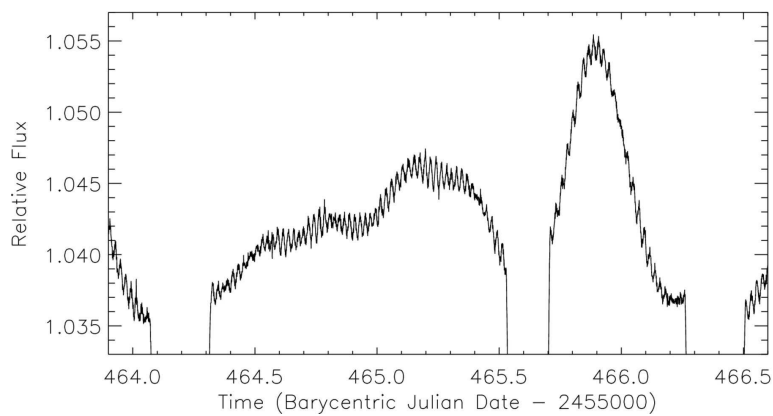}
\end{center}
\caption{Figure 2 from \citet{2013MNRAS.434..925H}.  An amplified image of the out-of-eclipse phase of the\,{\textit{Kepler}} Quarter 7 short-cadence light curve.  Both the $p$-mode (periods in the range 30 min to 1 h) and the $g$-mode pulsations ($\sim$1 d) are clearly visible. The pronounced periastron brightening can also be seen at approximately BJD 245 5466.}\label{hambletonfig2}
\end{figure}

One such\,{\textit{Kepler}} eclipsing heartbeat star is KIC 4544587, studied by \citet{2013MNRAS.434..925H}, with eccentricity 0.28, showing both high- and low-frequency modes typical of $\delta$ Sct and $\gamma$ Dor pulsations, as well as modes that are orbital frequency harmonics that may be excited by tidal resonances (See light curve excerpt in Fig. \ref{hambletonfig2}.) The masses of the two stars derived from binary modeling are 1.98 and 1.6 M$_{\odot}$.

\begin{figure}[h!]
\begin{center}
\includegraphics[width=8cm]{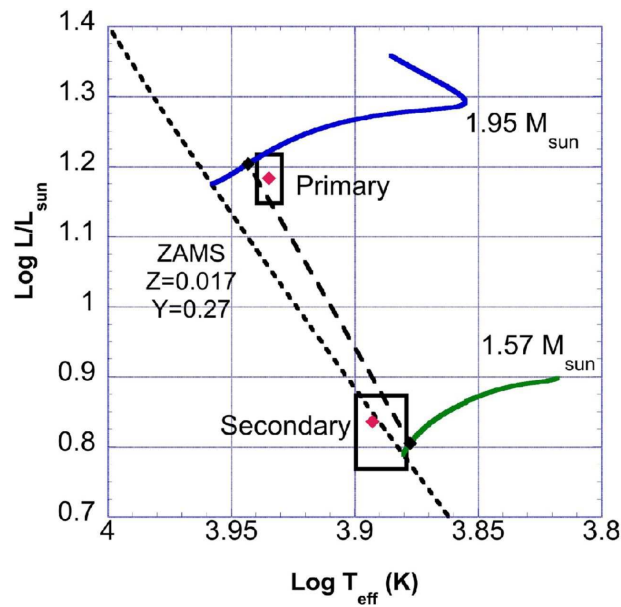}
\end{center}
\caption{Figure 13 from \citet{2013MNRAS.434..925H}.  An H-R diagram for stellar models of components of KIC 4544587. The boxes outline the parameter space for the observationally derived primary and secondary components. The short-dashed line is zero-age main-sequence position for stellar models with Z = 0.017, Y = 0.27. Also shown are evolutionary tracks for a 1.95-M$_{\odot}$ (blue) and 1.57-M$_{\odot}$ (green) model. The two models with the same age and composition closest to the observational constraints are connected by the long-dashed line. The red diamonds mark the best-fitting models for each star that do not have exactly the same age and composition.}
\label{hambletonfig13}
\end{figure}

An attempt was made to evolve models for each star with a common initial abundance and age, and using the same mixing length parameter, which best fit the constraints from binary orbits and pulsation modeling (Fig. \ref{hambletonfig13}). Pulsation calculations for the stars matching the derived constraints point to the 1.6 M$_{\odot}$ star as most likely being the $\delta$ Sct pulsator.  This exercise illustrated the potential of additional constraints from binaries to assist asteroseismic investigations. 

\vspace{0.1in}
\subsection{Amplitude variations}
\vspace{0.1in}

\begin{figure}[h!]
\begin{center}
\includegraphics[width=10cm]{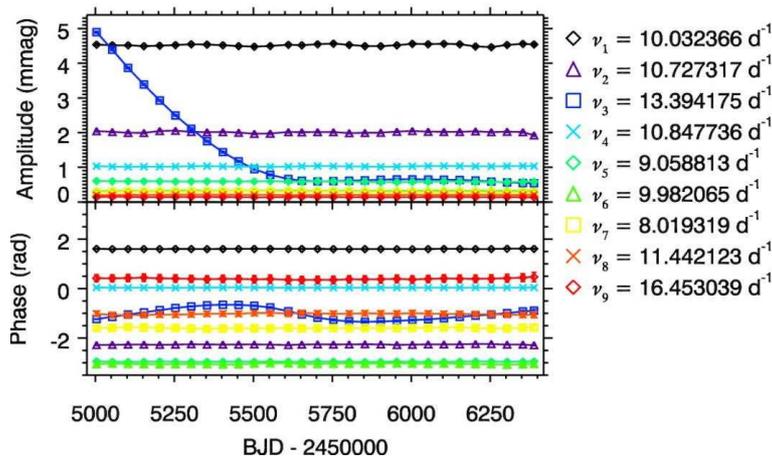}
\end{center}
\caption{Figure 8 from \citet{2016MNRAS.460.1970B}.  KIC 7106205 amplitude and phase tracking plot showing variability in pulsation mode amplitudes and phases over time.}\label{bowmanfig8}
\end{figure}

Most $\delta$ Sct asteroseismic studies have focused on periods and period spacings, but few have made use of the amplitudes of the pulsations.  The study of amplitudes requires nonlinear, nonradial, multidimensional hydrodynamic models which have not advanced far enough to predict mode selection and amplitudes of $\delta$ Sct stars.  While the frequency content, amplitudes, and phases of some $\delta$ Sct stars have been documented to change with time over many years, these phenomena were not investigated comprehensively until the project of \citet{2016MNRAS.460.1970B} using\,{\textit{Kepler}} data from 983 $\delta$ Sct stars observed continuously for four years.  Bowman et al. found that 61.3\% of stars in the sample showed amplitude variation in at least one pulsation mode during the four years.  One star, KIC 7106205, showed a remarkable amplitude decrease for a single frequency over the first two years of the\,{\textit{Kepler}} mission from 5 mmag to less than 1 mmag \citep{2014MNRAS.444.1909B} (Fig. \ref{bowmanfig8}).  The amplitude of this same mode was found from WASP data to have decreased from 11 mmag to 5 mmag during the two years prior to the\,{\textit{Kepler}} mission \citep{2015MNRAS.449.1004B}.  \citet{2014MNRAS.444.1909B} suggest that this dramatic decrease might be explained by nonlinear mode coupling with energy transfer from the $p$ mode to low-frequency high-degree $g$ modes that are not visible because their light variations average out over the stellar disk.  

\vspace{0.4in}
\subsection{Finding and interpreting frequency (period) spacing patterns—the key to asteroseismology}
\vspace{0.1in}

\begin{figure}[h!]
\begin{center}
\includegraphics[width=10cm]{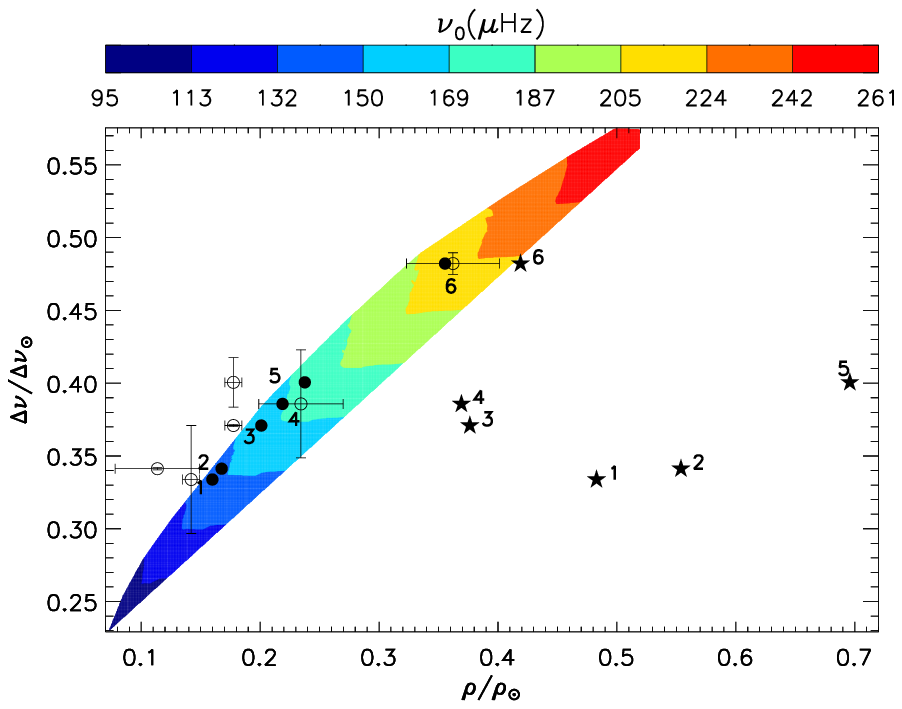}
\end{center}
\caption{Figure 2 from \citet{2014A&A...563A...7S}.  Predicted large separation as a function of the mean density of the star, normalized to their solar values 134.8 $\mu$Hz and 1.48 g cm$^{-3}$, respectively.  Color contours indicate the predicted frequency of the fundamental radial mode. Filled dots, empty dots, and star symbols represent mean densities found in \citet{2014A&A...563A...7S}, in the literature, and using the calibration of \citet{2011ApJ...726..112T}, respectively.  For the sake of clarity, the error bars in star symbol estimates are omitted, since they are larger than the abscissa range.  Reproduced with permission \copyright~ESO.} \label{suarezfig2}
\end{figure}

\begin{figure}[h!]
\begin{center}
\includegraphics[width=10cm]{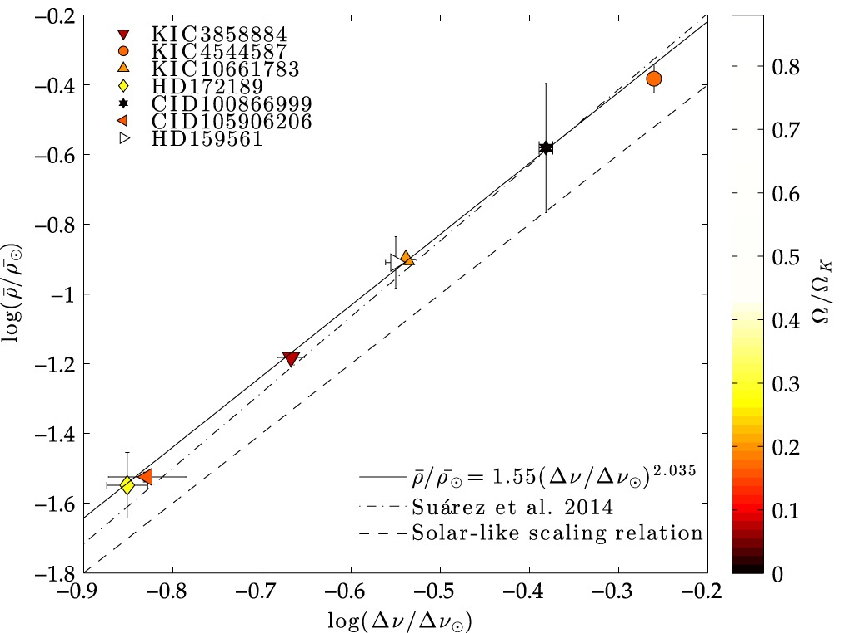}
\end{center}
\caption{Figure 1 from \citet{2015ApJ...811L..29G}.   Large separation--mean density relation obtained for seven binary systems. A linear fit to the points is also depicted, as well as the solar-like scaling relation from \citet{1980ApJS...43..469T}, and the theoretical scaling relation for non-rotating models of $\delta$ Sct stars from  \citet{2014A&A...563A...7S}. Symbols are plotted with a gradient colour scale to account for the different rotation rates.  \copyright~AAS. Reproduced with permission.} \label{garcia-hernandezfig1}
\end{figure}

Because $\delta$ Sct stars have modes of low radial order $n$, these modes cannot be treated using asymptotic pulsation theory and are not expected to show regular frequency spacings, such as the equal spacings between modes of consecutive radial order that are evident in, e.g., solar-like oscillators.  This lack of obvious regular frequency patterns, plus the more rapid rotation in these stars compared to the Sun, leading to asymmetric splitting of the modes into overlapping multiplets, has made it nearly impossible to identify the pulsation modes with certainty.  In addition, not all modes that are predicted by nonadiabatic pulsation calculations are seen in the amplitude spectrum.  Nevertheless, \citet{2014A&A...563A...7S}, using a grid of stellar models and calculating average frequency separations for degree $\ell$ =1 through 3 modes, show that an average large frequency spacing ($\Delta\nu$) can be determined and used to derive the mean stellar density (Fig. \ref{suarezfig2}).  \citet{2015ApJ...811L..29G} subsequently calibrated an observational frequency spacing – mean density relationship using eclipsing binaries with a $\delta$ Sct component observed by CoRoT and\,{\textit{Kepler}} to determine independently the mean density (Fig. \ref{garcia-hernandezfig1}).

\begin{figure}[h!]
\begin{center}
\includegraphics[width=10cm]{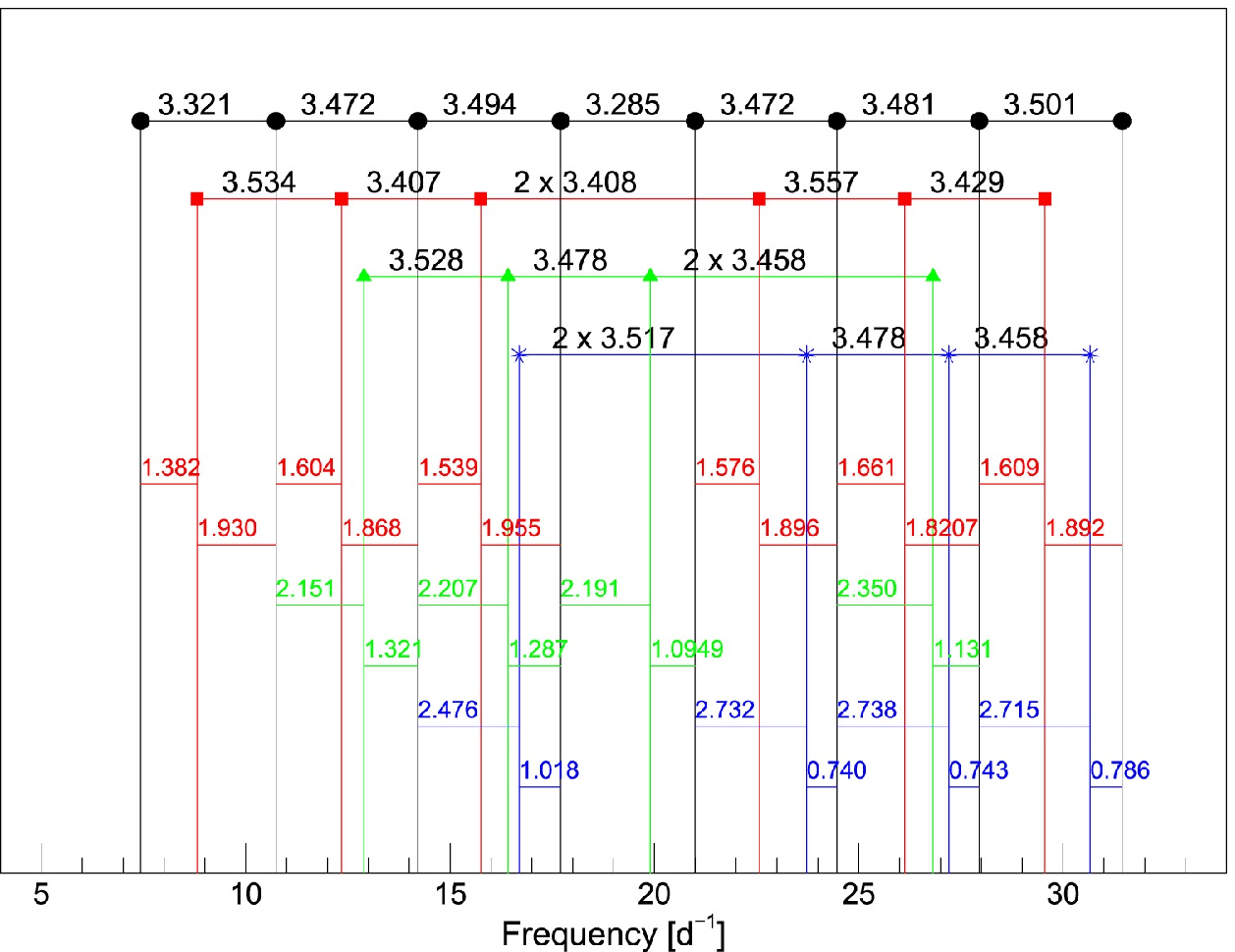}
\end{center}
\caption{Figure 1 from \citet{2016ApJ...822..100P}.  Sequences with quasi-equal spacing, and shifts of the sequences for CoRoT 102675756. First--black dots, average spacing 2.292$\pm$0.138 d$^{-1}$; Second--red squares, 2.290 $\pm$ 0.068 d$^{-1}$; Third--green triangles, 2.265$\pm$0.057 d$^{-1}$; Fourth--blue stars, 2.242$\pm$0.051 d$^{-1}$.  The mean spacing of the star is 2.277$\pm$0.088 d$^{-1}$. The shifts of the second, third, and fourth sequences relative to the first one are also given in the same color as the sequences.  \copyright~AAS. Reproduced with permission.} \label{paparo2016fig1}
\end{figure}

\begin{figure}[h!]
\begin{center}
\includegraphics[width=8cm]{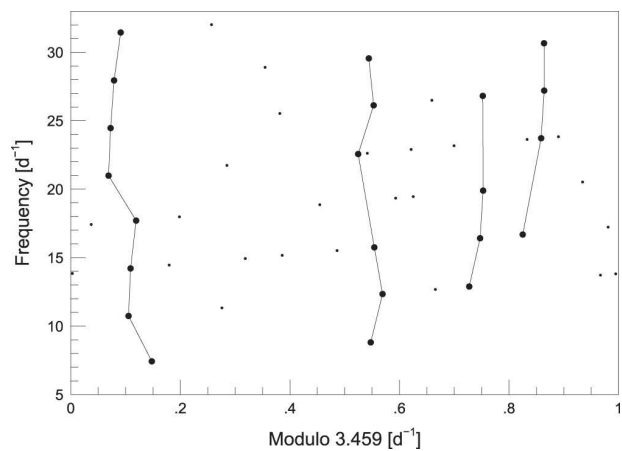}
\end{center}
\caption{Figure 2 from \citet{2016ApJ...822..100P}. Echelle diagram for CoRoT 102675756, consistent with the sequences of the Fig. \ref{paparo2016fig1} result from visual inspection. The mean spacing of the star was used as a modulo frequency. The whole frequency content of the star is plotted (small and large dots). The larger dots show the vertical representation of the sequences, the echelle ridges.  \copyright~AAS. Reproduced with permission.} \label{paparo2016fig2}
\end{figure}

\citet{2016ApJ...822..100P, 2016ApJS..224...41P} noticed by eye, and then confirmed by algorithm, that one or more sequences of characteristic spacings could be found in a sample of 90 $\delta$ Sct stars observed by CoRoT (Figs. \ref{paparo2016fig1} and \ref{paparo2016fig2}). It is not always easy to determine, however, whether these characteristic frequency spacings are between successive radial order modes of the same angular degree (i.e. represent $\Delta\nu$) or are instead a combination of $\Delta\nu$ and the rotational splitting frequency.

While these methods allow one to use $\delta$ Sct frequencies to determine a characteristic spacing and mean density, and point toward mode identifications, i.e., being able to identify sequences of modes of the same angular degree, they fall short of finishing the goal of detailed mode identification for asteroseismology.  

Fortunately, $\delta$ Sct stars early in their main-sequence lifetime have simpler patterns as their cores are less perturbed by changes in composition gradient at the convective core boundary that lead to `avoided crossings' and modes with mixed $p$- and $g$-mode character.  \citet{2020Natur.581..147B} found complete very regular sequences of $p$ modes among 60 young $\delta$ Sct variables observed by\,{\textit{Kepler}} and TESS (see Sec. \ref{sec:TESS}) and was able to use these to identify the modes, with an assumption that the highest amplitude mode in about 1/3 of the sample stars at frequency 18-23 d$^{-1}$ likely is the radial fundamental ($n$=1, $\ell$=0) $p$ mode, finally opening a window for $\delta$ Sct asteroseismology. This task may have been made easier by many stars in the sample having relatively slow rotation or possibly being observed pole-on, so that large rotational splittings did not confuse the sequence \citep{2020Natur.581..141B}.

\vspace{0.1in}
\subsection{$\gamma$ Doradus breakthroughs}
\vspace{0.1in}

A similar breakthrough for asteroseismology of $\gamma$ Doradus variables was enabled by the\,{\textit{Kepler}} data.  The $\gamma$ Dor $g$-mode periods are of high radial order, and should obey asymptotic period spacing relations.  These modes are sensitive to conditions at the convective core boundary,  where a composition gradient from hydrogen burning forms.  This gradient is altered by mixing from convective overshooting and differential rotation.  This composition gradient perturbs the expected even asymptotic period spacing, and causes mode trapping that shifts pulsation frequencies.  These deviations from even period spacing can be used to probe the region of the convective-core boundary, constrain mixing profiles, and measure interior rotation rates (See theoretical and modeling papers predicting and explaining these effects by \citet{2008MNRAS.386.1487M} and \citet{2013MNRAS.429.2500B}).

\begin{figure}[h!]
\begin{center}
\includegraphics[width=12cm]{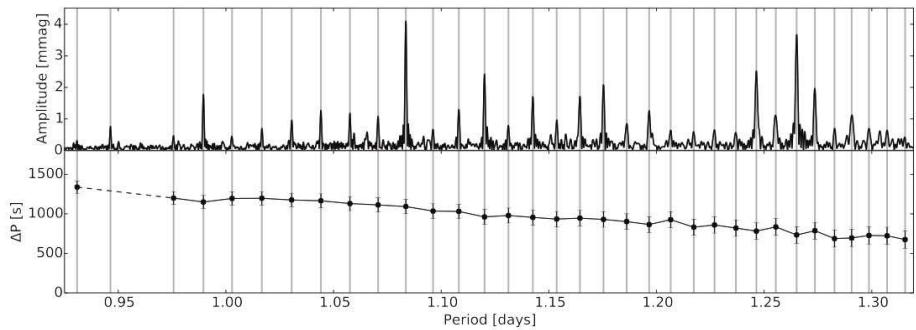}
\end{center}
\caption{Figure 11 from \citet{2015A&A...574A..17V}.  Top:  Close-up of part of the Fourier spectrum of KIC 11721304 (black). All the marked frequencies are accepted following the criterion described in \citet{2015A&A...574A..17V}.  Bottom: the period spacing as computed from the accepted frequencies. The black markers and grey lines indicate the frequencies for which a smooth spacing pattern was found.  Reproduced with permission \copyright~ESO.} \label{vanreeth2015fig11}
\end{figure} 

Despite the expectation that rapid rotation and overlapping non-equidistant rotationally split multiplets would make it difficult to find a period-spacing pattern, \citet{2015A&A...574A..17V, 2015ApJS..218...27V} were able to develop a period-spacing detection algorithm and successfully identify period spacings in several $\gamma$ Dor stars (Fig. \ref{vanreeth2015fig11}).  As is the situation for the $\delta$ Sct frequency spacings, this method by itself does not allow one to identify the radial order $n$, angular degree $\ell$ or azimuthal order $m$ of the modes.  Nevertheless, these techniques have been subsequently developed and applied for studies of $\gamma$ Dor stars observed by\,{\textit{Kepler}} and CoRoT (see, e.g., \citet{2017MNRAS.465.2294O, 2019MNRAS.482.1757L, 2019MNRAS.487..782L, 2020A&A...640A..49O}).

In the $\gamma$ Dor literature, the gravity or $g$ modes are also referred to as gravito-inertial modes, as they are subject to both buoyancy and Coriolis forces because of rotation in the star (see, e.g., \citet{2019MNRAS.485.3248M}).  A key recent development for $\gamma$ Dor stars was the detection and exploitation of the r modes, or global Rossby waves, (see, e.g., \citet{2018MNRAS.474.2774S}) in many $\gamma$ Dor stars observed by\,{\textit{Kepler}}.  These modes consist of predominantly toroidal motions that do not have a restoring force or cause light variation in a non-rotating star because they cause no compression or expansion. However, in a rotating star, the toroidal motion couples with spheroidal motion caused by the Coriolis force, resulting in temperature perturbations that are visible. In $\gamma$ Dor stars, the r modes of a given azimuthal order $m$ appear in groups with slightly lower frequency than $m$ times the rotation frequency.  These modes can be excited by the $\kappa$ mechanism.

Gravity mode and inertial (r) mode period spacing patterns found in\,{\textit{Kepler}} $\gamma$ Dor stars, in some cases supplemented by $\delta$ Sct $p$ modes for the hybrid pulsators, have since been used by many groups to determine core and envelope rotation profiles in these stars (see, e.g., \citet{2016A&A...593A.120V, 2018A&A...618A..24V, 2017MNRAS.465.2294O, 2019MNRAS.487..782L, 2020MNRAS.491.3586L}).  Interior differential rotation has also been studied and confirmed in $\gamma$ Dor stars using the mode spacing patterns by, e.g., \citet{2017ApJ...847L...7A, 2018A&A...618A..24V, 2018A&A...618A..47C}. 

Many important modeling advances for $\gamma$ Dor stars have resulted from the\,{\textit{Kepler}} observations, among these models of individual targets (e.g., \citet{2014MNRAS.444..102K, 2015MNRAS.447.3264S, 2016A&A...592A.116S}).  \citet{2019MNRAS.485.3248M,2020ApJ...895...51M} have developed comprehensive model grids to interpret mode spacing patterns for ensembles of $\gamma$ Dor stars and evaluate the effects of including diffusive settling and radiative levitation.

\vspace{0.1in}
\subsection{$\delta$ Sct discoveries in binaries and clusters}
\vspace{0.1in}

While not discussed in detail here, the\,{\textit{Kepler}} mission discovered many eclipsing binary systems.  As mentioned earlier, the stellar properties derived independently from the binary system modeling can be used to calibrate asteroseismic techniques and supplement constraints for detailed asteroseismic models.  A catalog of such systems discovered or observed by\,{\textit{Kepler}} and K2 is maintained by Villanova University, with links to the Mikulski Archive for Space Telescopes (MAST), and the VizieR On-line Data Catalog (see, e.g., \citet{2016AJ....151...68K, 2016yCat..51510068K}.)  As of this writing, the Third Edition of this catalog lists 2922 binaries observed during the original\,{\textit{Kepler}} mission\footnote{http://keplerebs.villanova.edu} and 664 binaries from K2\footnote{http://keplerebs.villanova.edu/k2}.

Many of these binaries have potential $\delta$ Sct (or hybrid) components.  For example, \citet{2016arXiv160608638L, 2017MNRAS.465.1181L} catalog properties of 199 systems, many from\,{\textit{Kepler}} observations, and \citet{2019A&A...630A.106G} find 149 such systems in the\,{\textit{Kepler}} data.  See also \citet{2017MNRAS.470..915K} who discuss spectroscopic observations of 92 eclipsing binaries with a $\delta$ Sct component apart from the\,{\textit{Kepler}} data.  We can add to this list 341\,{\textit{Kepler}} non-eclipsing A/F star binaries discovered from pulsation phase modulation by \citet{2018MNRAS.474.4322M, 2018yCat..74744322M}, and the 176\,{\textit{Kepler} heartbeat stars mentioned earlier \citep{2016AJ....151...68K}.  There are a few dozen \,{\textit{Kepler}} binary systems containing a $\delta$ Sct component that have been studied in detail to date.  A search on the ADS abstract service with terms 'binary' and `$\delta$ Sct' in the abstract, and `KIC' (Kepler Input Catalog) in the title gives 42 results, and shows a trend of  increasing number of publications from two to ten per year over the past decade.  Searching for publications on binary\,{\textit{Kepler}} objects with `$\gamma$ Dor' in the abstract yields 28 results, with many of the objects studied being hybrid pulsators.  This search does not include papers that are likely to appear soon studying binaries observed during the K2 mission.

$\delta$ Sct stars found in clusters have similar advantages to those in binaries, as the common age and metallicity of the cluster members offer independent constraints.  For example, a comprehensive paper by \citet{2016ApJ...831...11S} compares the age and distance modulus of the open cluster NGC 6811 derived using a variety of methods.  This cluster was observed during the original\,{\textit{Kepler}} mission, and found to contain an eclipsing binary with an Am star and $\gamma$ Dor component, many pulsating stars near the cluster turnoff, including 28 $\delta$ Sct, 15 $\gamma$ Dor, and 5 hybrid stars, and many giant stars, some on the asymptotic giant branch and some in the 'red clump' core helium-burning phase.  They model the binary to determine component masses and compare with isochrones, finding inconsistent ages between the components of 1.05 Gyr for the Am star primary, and 1.21 Gyr for the $\gamma$ Dor secondary; the younger Am star age is more consistent with the 1.0 Gyr age derived from the main-sequence turnoff in the color-magnitude diagram.  However, the Am star abundance peculiarities are not properly taken into account in stellar models, so the Am star age may be more suspect.  In addition, the (near-solar) metallicity of the stars in the cluster is uncertain and the ages would be less discrepant with a slightly lower metallicity than adopted.  Applying asteroseismic $\Delta\nu$ vs. $\nu_{\rm max}$ relations for the core helium-burning stars to determine their radii and masses, they find that the derived masses of these stars appear to be larger (or their radii smaller) than expected for the cluster age(s) derived from the binary. These stars are more consistent with a 0.9 Gyr age.  They derive the distance modulus using the period-luminosity relationship of high-amplitude $\delta$ Sct stars, finding ($m$-$M$)$_{V}$ = 10.37 $\pm$ 0.03, which is lower than the value derived using the eclipsing binary, ($m$-$M$)$_{V}$ = 10.47 $\pm$ 0.05.  This example shows the possibilities for combining multiple constraints from clusters to check for consistency in inferences, and to identify discrepancies in modeling or asteroseismic analysis techniques for a particular star class.

\section{Successes, unresolved problems and questions}

In summary, it would not be an understatement to conclude that the\,{\textit{Kepler}}  mission has revolutionized the field of asteroseismology, in particular for $\delta$ Sct stars, and given researchers a wealth of data for analyses, modeling efforts, and motivating future long-term observations.  A non-exhaustive list of successes includes:

\begin{itemize}
\item Unprecedented long time-series (months to years), high-cadence (1 min or 30 min), high-precision photometry for thousands of $\delta$ Sct, $\gamma$ Dor, and hybrid variables, many newly discovered using the\,{\textit{Kepler}} data.
\item Interpretation and application of frequency and period spacings and patterns to inform mode identification, exploiting these to determine interior structure, extent of mixing, and rotation profiles.
\item More definitive quantification of the pervasiveness and range of amplitude variations that await explanation.
\item Large expansion in the number of binaries showing tidally excited modes, establishing a new class of variable stars.
\item Motivation for exploration of additional pulsation driving mechanisms, advancing stellar pulsation theory.
\item More definitive data to quantify the role of magnetic fields in pulsation.
\item More definitive constraints for quantifying the effects of element diffusive settling and radiative levitation and accurately including these processes in stellar models.
\end{itemize}

There are many problems and questions motivated or amplified by the\,{\textit{Kepler}} data, among these:

\begin{itemize}
\item Why are many of the pulsation modes predicted by linear pulsation theory not observed?
\item What is the origin of the low frequencies found in many $\delta$ Sct stars?
\item Can observed frequency spacings and patterns be interpreted and used for mode identification?
\item What determines the amplitudes of $\delta$ Sct modes, and what causes amplitude variations? 
\item Why are some stars in the $\delta$ Sct (and $\gamma$ Dor) instability regions ‘constant’, i.e. not pulsating?
\item Why are some chemically peculiar stars pulsating?
\item Are HADS or SX Phe stars different from each other or from normal $\delta$ Sct stars?
\item What is the origin of blue stragglers?
\item What is the origin of magnetic activity, spots, and flares in hot stars?
\item What is the origin of the abundance peculiarities in $\lambda$ Boo, Am and Ap stars?
\item Can new proposed pulsation driving mechanisms explain the unexpected frequencies observed in some $\delta$ Sct stars?
\end{itemize}

Answering these questions will require long-term monitoring, directed campaigns, high-resolution spectroscopy, multicolor photometry, interferometry, and other observations, in addition to advances in stellar evolution and pulsation theory and modeling.

\section{The near future and TESS}
\label{sec:TESS}
While the\,{\textit{Kepler}} spacecraft ended its K2 mission in November 2018, the NASA TESS spacecraft \citep{2015JATIS...1a4003R} was launched in April 2018. The TESS spacecraft has some advantages and some disadvantages compared to\,{\textit{Kepler}} for asteroseismology.  The TESS mission is surveying more of the sky over its mission lifetime, while\,{\textit{Kepler}} covered a single field of view in the Cygnus and Lyra constellations during its original mission, and 18 fields along the ecliptic during the extended K2 mission.  However, TESS observes a sector of the sky continuously for only 27 days, compared to the possibility of obtaining up to four years of nearly continuous data during the original\,{\textit{Kepler}}  mission, or nearly three months continuously during K2.  TESS is collecting full-frame images every 30 minutes (every 10 minutes starting in Cycle 3), and also has the possibility for 2-minute and even 20-second cadence observations for selected targets.  The pixel size for the TESS cameras is larger, making crowding and contamination from nearby stars in the field an issue that must be taken into account in data analyses.  The redder TESS bandpass reduces the observed amplitudes of $\delta$ Sct pulsations by about 25 percent compared to the amplitudes of the\,{\textit{Kepler}} mission \citep{2019MNRAS.490.4040A}.  

The TESS first-light papers have been published, including a first view of $\delta$ Sct and $\gamma$ Dor stars with the TESS mission \citep{2019MNRAS.490.4040A}. This paper contains up-to-date descriptions of $\delta$ Sct and related stars, including pre-main-sequence $\delta$ Sct stars that were not studied in detail using\,{\textit{Kepler}} observations, $\lambda$ Boo stars not discussed in this review, TESS observations of very bright stars such as $\alpha$ Pic, and the pulsation class prototypes SX Phe (see also \citet{2020svos.conf...81D,2020MNRAS.499.3034D}) and $\gamma$ Dor that were not targeted by\,{\textit{Kepler}}.  The paper also has an extensive explanation of the role of turbulent pressure in the hydrogen ionization zone in driving $\delta$ Sct pulsations, especially in the context of Am stars that are expected to have helium depleted from diffusive settling, inhibiting the classical $\kappa$-effect pulsation driving mechanism.  TESS observations will extend and enhance the\,{\textit{Kepler}} legacy.

What can we expect for the future of\,{\textit{Kepler}} and TESS observations?  Asteroseismic analyses will be conducted using data from individual stars or ensembles of stars with common properties, binaries, and $\delta$ Sct stars in clusters.  K2 has observed many open clusters on the ecliptic that contain $\delta$ Sct stars (e.g., Hyades, Praesepe, Pleiades, and M67).  Studies of clusters show promise to finally understand the nature of blue stragglers, and the development of abundance peculiarities in Am and $\lambda$ Boo stars.  There were no pre-main-sequence $\delta$ Sct stars in the original\,{\textit{Kepler}} field, but discoveries for these stars may await using K2 or TESS data.  It is hoped that these data will help disentangle or systematize the picture for stellar interior and evolution modeling from the pre-main-sequence through the shell H-burning stage, for example, the roles of processes such as convective overshooting, differential rotation, angular momentum transport, element levitation and settling, magnetic fields, mixing from internal gravity waves, etc.  It is hoped that advances in theory and multidimensional stellar modeling, e.g., nonradial, nonlinear, nonadiabatic pulsation modeling including turbulent and magnetic pressure and energy or rapid differential rotation, will lead to explanations for pulsation mode driving, mode selection, and amplitudes, and will better define instability strip boundaries.  

\section*{Author Contributions}

J.G. is the sole author of this article that reviews the contributions of many authors to the field of $\delta$ Sct asteroseismology using\,{\textit{Kepler}} data.

\section*{Funding}
J. Guzik's research is supported at Los Alamos National Laboratory (LANL), managed by Triad National Security, LLC for the U.S. Department of Energy's NNSA, Contract \#89233218CNA000001. J.G. also gratefully acknowledges a Los Alamos National Laboratory Center for Space and Earth Sciences Rapid Response grant for Summer 2020.

\section*{Acknowledgments}
J.G. thanks editors K. Kinemuchi and A. Baran for the opportunity to write this review, and for their careful reading and suggestions. J.G. also thanks many colleagues who have made important contributions to this field. In addition. J.G. thanks the two reviewers, and also Simon Murphy, who have provided extensive suggestions for additional literature to take into account, and for sharing their perspectives on the developments in this field.  J.G. thanks colleagues J. Jackiewicz and P. Bradley for advice and help with the\,{\textit{Kepler}} data over many years.  Finally, J.G. thanks the authors and journals for granting permission to reproduce the figures used in this paper.

\bibliographystyle{frontiersinSCNS_ENG_HUMS} 
\bibliography{GuzikFrRev.bib}



\end{document}